\documentstyle[buckow]{article}
\newcommand{\half}{{1 \over 2}}
\newcommand{\dg}{{\dagger}}
\newcommand{\p}{\partial}
\newcommand{\Tr}{{\rm Tr}\,}
\newcommand{\CD}{{\cal D}}
\newcommand{\CN}{{\cal N}}

\newcommand{\CZ}{{\cal Z}}
\newcommand{\CO}{{\cal O}}
\newcommand{\CS}{{\cal S}}

\newcommand{\CL}{{\cal L}}
\newcommand{\CI}{{\cal I}}

\begin {document}
\begin{flushright}
{\small\hfill AEI-106\\
\hfill hep-th/9904111}\\
\end{flushright}
\def\titleline{ 
%%%%%%%%%%%%%%%%%%%%%%%%%%%%%%%%%%%%%%%%%%%%%%
%                                            %
% Insert now the text of your title.         %
% Make a linebreak in the title with         %
%                                            %
%            \newtitleline                   %
%                                            %
%%%%%%%%%%%%%%%%%%%%%%%%%%%%%%%%%%%%%%%%%%%%%%
Large $N$ Matrix Field Theories\footnote{
Talk given at the 32nd International Symposium Ahrenshoop on the Theory of 
Elementary Particles, Buckow, Germany, 1-5 Sep 1998.} 
%\newtitleline
%title, continued
%%%%%%%%%%%%%%%%%%%%%%%%%%%%%%%%%%%%%%%%%%%%%%
}
\def\authors{
%%%%%%%%%%%%%%%%%%%%%%%%%%%%%%%%%%%%%%%%%%%%%%
%                                            %
%  Insert now the name (names) of the author %
%  (authors).                                %
%  In the case of several authors with       %
%  different addresses use e.g.              %
%                                            %
%             \1ad  , \2ad  etc.             %
%                                            %
%  to indicate that a given author has the   %
%  address number 1 , 2 , etc.               %
%                                            %
%%%%%%%%%%%%%%%%%%%%%%%%%%%%%%%%%%%%%%%%%%%%%%
Matthias Staudacher
%%%%%%%%%%%%%%%%%%%%%%%%%%%%%%%%%%%%%%%%%%%%%
}
\def\addresses{
%%%%%%%%%%%%%%%%%%%%%%%%%%%%%%%%%%%%%%%%%%%%%%
%                                            %
% List now the address. In the case of       %
% several addresses list them by numbers     %
% using e.g.                                 %
%                                            %
%             \1ad , \2ad   etc.             %
%                                            %
% to numerate address 1 , 2 , etc.           %
%                                            %
%%%%%%%%%%%%%%%%%%%%%%%%%%%%%%%%%%%%%%%%%%%%%%
Albert-Einstein-Institut, Max-Planck-Institut f\"{u}r
Gravitationsphysik\\ Schlaatzweg 1\\  D-14473 Potsdam, Germany
%%%%%%%%%%%%%%%%%%%%%%%%%%%%%%%%%%%%%%%%%%%%%%%
}
\def\abstracttext{
%%%%%%%%%%%%%%%%%%%%%%%%%%%%%%%%%%%%%%%%%%%%%%%
%                                             %
% Insert now the text of your abstract.       %
%                                             %
%%%%%%%%%%%%%%%%%%%%%%%%%%%%%%%%%%%%%%%%%%%%%%%
We discuss aspects of recent novel approaches towards understanding the large
$N$ limit of matrix field theories with local or global non-abelian symmetry.
%%%%%%%%%%%%%%%%%%%%%%%%%%%%%%%%%%%%%%%%%%%%%%%%%%%%%%%%%%%%%%%
}
\large
\makefront
%%%%%%%%%%%%%%%%%%%%%%%%%%%%%%%%%%%%%%%%%%%%%%%%
%                                              %
%  Insert now the remaining parts of           %
%  your article.                               %
%                                              %
%%%%%%%%%%%%%%%%%%%%%%%%%%%%%%%%%%%%%%%%%%%%%%%%
% 4 resp. 6 pages!
The large $N$ limit of field theories whose local degrees of freedom
are $N \times N$ matrices continues to intrigue physicists since its
discovery by `t Hooft 25 years ago \cite{thooft}. 
This is striking given the fact that
no field theory with propagating matrix degrees of freedom has up to now been 
exactly solved by large $N$ techniques. Certainly, one reason for this 
continuing interest is the
significant evidence that this limit indeed entails important simplifications.
Furthermore, in the most interesting cases, physics appears to be not
too different as compared to the finite $N$ case. Lastly, a new aspect of
``large $N$'' has appeared over the years: $N=\infty$ may not only be
a reasonable and tractable {\it approximation} to some theory, but
may also {\it define the theory}. Indeed, we know this to be true 
in some simple cases such as non-critical string theory, and now
there are serious proposals for the case of ``real'' (i.e.~critical,
supersymmetric) string theory and 11-dimensional M-theory.

One of the above mentioned simplifications that seems to occur can be
roughly summarized by the correspondence
\begin{equation}
\begin{array}{ccc}
{\rm matrix~field~theory}\qquad &\simeq & \qquad  
{\rm~suitable~zero-dimensional~matrix~model} \cr
(N=\infty) &  & (N=\infty)
\end{array}
\label{equi}
\end{equation}
That is, it appears that every matrix field theory can be replaced by
a suitable matrix integral such that at least some of the physical 
observables on both sides become identical at $N=\infty$. A first 
instance of this was discovered by Eguchi and Kawai \cite{ek},
but we now know quite a few further examples, and a much more general
principle, as loosely stated in the equivalence (\ref{equi}),  
appears to be at work. 

In part 1 we will discuss a naive (presumably too naive) version
of the correspondence (\ref{equi}) for Yang-Mills field theory, where
we point out that a surprisingly simple reduced matrix model, contrary
to initial expectations, proves to be well-defined for large $N$. 
Our main point
here is that the existence of the proposed Yang-Mills matrix integrals has
been overlooked in the past, and that they are, apart from other
important applications, an ideal
laboratory for testing new large $N$ gauge theory techniques.  
Part 2 deals with known exact equivalences (\ref{equi}) for
interacting $D$-dimensional lattice field theories with global
$U(N)$ symmetry, and we outline, taking as a specific example a $D=2$
hermitian model, a very general procedure for bootstrapping the 
$N=\infty$ solution.

\section{Local $U(N)$: Yang-Mills integrals}
{\it In this part we discuss some aspects of results
obtained in collaboration with W.~Krauth and H.~Nicolai and
published in} \cite{kns},\cite{ks1},\cite{ks2}.
Consider $D$-dimensional pure $SU(N)$
Yang-Mills field theory and, inspired by
the principle (\ref{equi}), reduce it by brute force to zero dimensions.
The continuum path integral, involving traceless hermitian gauge connections
$X_\mu$, becomes an ordinary matrix integral:
\begin{equation}
\CZ_{D,N} = \int \prod_{A=1}^{N^2-1} \prod_{\mu=1}^{D}  
\frac{d X_{\mu}^{A}}{\sqrt{2 \pi}}
\exp \bigg[  \frac{1}{2} \Tr
[X_\mu,X_\nu] [X_\mu,X_\nu] \bigg].
\label{bosint}
\end{equation}
Note that gauge fixing is no longer required here, since the overcounting
of gauge-equivalent configurations involves merely a factor of the compact,
finite volume of the gauge group: space time has become a point (or more
precisely, an infinitesimal torus, since the ``point'' still keeps a sense
of the $D$ directions.). Now, as was explained in \cite{gk}, the integral
eq.(\ref{bosint}) still ``knows'' something about $D$-dimensional 
space-time. Indeed, shifting
\begin{equation}
X_\mu \rightarrow P_\mu + X_\mu
\label{shift}
\end{equation}
by diagonal matrices $P_\mu=$diag$(p^1_\mu, \ldots, p^N_\mu)$ 
we formally recover Feynman rules which
look like the ordinary ones except that the momentum integrations are
replaced by sums over discretized momenta $p^i_\mu-p^j_\mu$. As
$N \rightarrow \infty$ one might hope that the sums turn back into
loop integrals, motivating the correspondence (\ref{equi}).
Now in \cite{gk} a somewhat complicated quenching and gauge fixing procedure 
was introduced in order to ensure the recovery of the field theory.
Indeed it would seem at first sight that the integral eq.(\ref{bosint}) is
meaningless without the procedure of \cite{gk} since there are unconstrained
flat directions in integration space, due to mutually commuting matrices.
However, the Monte Carlo results of \cite{ks1} suggest

\vspace{0.5cm}\noindent
{\bf Proposition 1a:} The Yang-Mills 
integrals $\CZ_{D,N}$ exist iff $N> {D \over D-2}$.
\vspace{0.5cm}

\noindent
It would be quite important to find methods enabling one to rigorously
prove this statement, or even calculate the partition sums $\CZ_{D,N}$. 
Some important analytic evidence comes from the
perturbative calculations of \cite{nishi}. For $SU(2)$, a proof of
the proposition, as well as an analytic expression for $\CZ_{D,2}$, is
known.

The matrix integrals eq.(\ref{bosint}) have beautiful supersymmetric
extensions in dimensions $D=4,6,10$. These read
\begin{equation}
\CZ_{D,N}^{\CN}:=\int \prod_{A=1}^{N^2-1} 
\Bigg( \prod_{\mu=1}^{D} \frac{d X_{\mu}^{A}}{\sqrt{2 \pi}} \Bigg) 
\Bigg( \prod_{\alpha=1}^{\CN} d\Psi_{\alpha}^{A} \Bigg)
\exp \bigg[  \frac{1}{2} \Tr 
[X_\mu,X_\nu] [X_\mu,X_\nu] + 
\Tr \Psi_{\alpha} [ \Gamma_{\alpha \beta}^{\mu} X_{\mu},\Psi_{\beta}]
\bigg].
\label{susyint}
\end{equation}
where we have supersymmetrically 
added $\CN=2 (D-2)$ hermitian fermionic matrices 
$\Psi_{\alpha}$ to the models. The $D=10$ model corresponds to the
dimensional reduction of the maximally supersymmetric conformal
$D=4,\CN=4$ Yang-Mills field theory to zero dimensions. It is
also the crucial ingredient in the IKKT model for IIB superstrings
\cite{ikkt1},
which however, instead of taking the large $N$ limit, sums 
$\CZ_{10,N}^{16}$ over all
values of $N$. Following the $SU(2)$ calculations of \cite{sestern},
the perturbative arguments of \cite{ikkt2}, the arguments of \cite{greengut},
the calculations of \cite{moore}, and our Monte Carlo work, we are led 
to 

\vspace{0.5cm}\noindent
{\bf Proposition 1b:} The susy Yang-Mills 
integrals $\CZ_{4,N}^4,\CZ_{6,N}^8,\CZ_{10,N}^{16}$ exist iff $N \geq 2$.
\vspace{0.5cm}

\noindent
The analytic results of these integrals are believed to be known, and
a rigorous mathematical proof would be welcome.

It is interesting to understand the similarities and differences of these
little studied
``{\it new}'' matrix models eqs.(\ref{bosint}),(\ref{susyint}), whose
existence has been missed until recently, in relation to
the conventional ``{\it old}''
matrix models of Wigner type. A crucial quantity in the old matrix models
is the distribution of eigenvalues of the random matrices. An interesting
novel feature of the new matrix models is the fact that, at finite $N$,
only a finite number of one-matrix moments exist. The numerical
results agree with perturbative powercounting arguments, and one
is led, for the bosonic models eq.(\ref{bosint}), to

\vspace{0.5cm}\noindent
{\bf Proposition 2a:}  
$\Big\langle {1 \over N}\Tr X_1^{2 k} \Big\rangle < \infty$ iff 
$k<N (D-2) - {3\over 2} D+2$,
\vspace{0.5cm}

\noindent
while in the supersymmetric cases $D=4,6,10$  eq.(\ref{susyint}) one has 

\vspace{0.5cm}\noindent
{\bf Proposition 2b:}  
$\Big\langle {1 \over N}\Tr X_1^{2 k} \Big\rangle < \infty$ iff 
$k<D-3$.
\vspace{0.5cm}

\noindent
Once again, except for $SU(2)$, rigorous proofs of these conjectures
are missing.
These findings indicate 
that in the new matrix models the density of eigenvalues 
falls off much slower (powerlike) than in the old ones (exponential).
As $N \rightarrow \infty$ the bosonic densities behave once again rather
conservatively (infinitely many moments exist), while for the susy densities
the behavior indicated in proposition 2b is {\it independent} of $N$.

A much more difficult question is whether these models might lead
to a ``self-quenching'' effect where a background $P_\mu$ 
(in eq.(\ref{shift})), bearing
some resemblance to real Yang-Mills theory, 
is dynamically generated as $N \rightarrow \infty$.

The above Yang-Mills integrals have many applications even at finite
$N$ (for a recent unexpected one see \cite{hollo}); however, here
we would like to stress that they constitute an ideal laboratory
for developing new large $N$ techniques aimed at making progress
with `t~Hooft's large $N$ QCD \cite{thooft}.

\section{Global $U(N)$: Master partitions}

The problem of finding the $N=\infty$ solution to matrix field theories
has not even been solved in the presumably simpler case of models
with a global $U(N)$ symmetry. The main obstacle has been that 
no systematic procedure was known to reduce the local number of degrees
of freedom from $\CO (N^2)$ to $\CO (N)$. In \cite{master} we outlined
a general approach for achieving such a reduction for any
field theory with a global matrix symmetry. Let us sketch the idea
in the specific example of an interacting $D=2$ hermitian 
scalar field theory. It is convenient to put the theory on a lattice:
\begin{displaymath}
\CZ= \int \prod_x \CD M(x)~e^{ -\CS},
\end{displaymath}
\begin{equation}
\CS= N \Tr \sum_x \bigg[ \half M(x)^2 + {g \over 4} M(x)^4 - 
 {\beta \over 2} \sum_{\mu=1,2}
[M(x) M(x+\hat{\mu}) + M(x) M(x-\hat{\mu})]  \bigg],
\label{mlattice}
\end{equation}  
where the field variables are $N\times N$ hermitian matrices $M(x)$
defined on the square lattice sites $x$ and $\hat{\mu}$ 
denotes the unit vector in the $\mu$-direction. The measure is the
usual flat measure on hermitian matrices. The first step consists in
applying the reduction principle (\ref{equi}). Naively reducing the system
as in the previous section
down to a single 
point results in an ordinary one-matrix model where the information
on the 2$D$ lattice is lost. A more careful reduction has to hide the
propagation on the lattice in group space; here we will use the
beautiful procedure of ``twisting'', see \cite{twisted} and references 
therein. Using the $N \times N$ Weyl-`t~Hooft matrices
\begin{equation}
P=\left( \begin{array}{cccccc}
0 & 1 &   &  &  &  \\
  & 0 & 1 &  &  &   \\
  &   & \ddots & \ddots &  &  \\
  &   &        &        & 0  & 1\\
1 &  & & &  & 0
\end{array} \right),
\quad Q=\left( \begin{array}{cccccc}
1 &  &   &  &  &  \\
  & \omega &   &  &  &   \\
  &   & \ddots &  &  &  \\
  &   &        &        & \omega^{N-2}  & \\
  &  & & &  & \omega^{N-1}
\end{array} \right),
\label{pq}
\end{equation}
where $\omega=\exp {2 \pi i \over N}$ and $P Q=\omega Q P$, one can show
by a Fourier transform in matrix index space that the one-matrix integral
\begin{equation}
Z= \int \CD M \exp 
N \Tr \bigg[
-\half M^2 - {g \over 4} M^4 
+\beta \Big( M P M P^\dg + M Q M Q^\dg \Big)  \bigg],
\label{mtwist}
\end{equation}  
has the same vacuum energy as the path integral eq.(\ref{mlattice}).

As a second step we need to reduce the number of variables from $N^2$
to $N$. The brute force approach would be to diagonalize the matrix
$M$ and perform the integration over the unitary diagonalizing matrix.
One would then obtain an effective action for the $N$ eigenvalues of
$M$. However, calculations at small $N$ show that this effective action
is extremely complicated in the case at hand. On the other hand, 
if we change variables
from the eigenvalues to {\it partitions}, corresponding to a Fourier
transform in group space, something very interesting happens. The $N$
variables dual to the $N$ 
eigenvalues are the Young weights $h_i=N-i+m_i$, $i=1, \ldots, N$,
where the $m_i$ are the lengths of the $i$'th row in the Young diagram
corresponding to the partition. Denoting the partitions
through $h=(h_1, \ldots, h_N)$, the dual representation of the integral
eq.(\ref{mtwist}) is found to be
\begin{equation}
Z=\sum_h \CI_h~\CL_h~\beta^{{|h| \over 2}},
\label{mfull}
\end{equation}
where instead of an integration over the $N\times N$ matrix $M$ we
now have a sum over all partitions $h$ of the non-negative integer
$|h|=0,1,2,\ldots$. Here $\CI_h$ contains all the information
on the interaction, and essentially requires the general correlation
function of the $U(N)$-invariant one-matrix integral
\begin{equation}
\CI_h=N^{|h|} \prod_{i=1}^N {(N-i)! \over h_i!}~
\int \CD M~\exp~N \Tr \Big[ -\half M^2 -{g \over 4} M^4 \Big]~\chi_h(M),
\label{hermitian}
\end{equation}
which are known.
Here $\chi_h(M)$ are the Schur functions on $h$ which are nothing but
a complete set of of class functions (non-abelian Fourier modes) 
on the group.
The information on the lattice is contained in the
{\it lattice polynomials} 
\begin{equation}
\CL_h=\exp{1 \over N} \Tr \Big(\p P \p P^\dg + \p Q \p Q^\dg \Big) 
\cdot \chi_h(J)~\Big|_{J=0}.
\label{poly}
\end{equation}
Here $\p$ denotes the $N \times N$ matrix differential operator
whose matrix elements are $\p_{j i}={\p \over \p J_{i j}}$.
The $\CL_h$ are easily shown to be polynomials in the
variable ${1 \over N}$ of maximal degree $\half |h| -1$.

Now the result of this harmonic analysis 
is that the terms to be summed over partitions in 
eq.(\ref{mfull}) {\it factorize} into a piece $\CI_h$ containing the
information on the local interaction and and the piece $\CL_h$ containing
the information on the space-time structure. Since there are only $N$
variables $h_i$ we expect the sum eq.(\ref{mfull}) to be dominated at
$N=\infty$ by a saddle point, i.e.~an effective master partition.
In a third and final step 
we will need to write the full system of bootstrap equations
for the saddle point. This will require a deeper analysis of the lattice 
polynomials. But it should be clear that the problem of solving the
large $N$ lattice field theory has been reformulated in a rather
non-trivial way: In fact, the interacting theory (i.e. $g \neq 0$ in 
eq.(\ref{mlattice})) is no harder to solve in this dual
space of Young weights than the free theory ($g=0$).

\vskip0.5cm
\noindent
{\large \bf Acknowledgements}

\smallskip
\noindent
I thank W.~Krauth and H.~Nicolai for fruitful collaboration,
and J.~Hoppe, V.~A.~Kazakov, I.~K.~Kostov,
and J.~Plefka for useful discussions.
This work was supported in part by the EU under Contract FMRX-CT96-0012.

%%%%%%%%%%%%%%%%%%%%%%
%%%%%%%%%%%%%%%%%%%%%%

\end{document}